\documentclass[aps,prd,twocolumn,superscriptaddress]{revtex4-1}
\usepackage{graphicx}
\usepackage{amsfonts}
\usepackage{amsmath}
\usepackage{comment}
\usepackage{subcaption}
\usepackage{siunitx}
\usepackage{float}
\usepackage{hyperref}
\usepackage{placeins}
\usepackage[all]{hypcap}
\hypersetup{
colorlinks=true,
linkcolor=blue,
citecolor=blue,
urlcolor=blue
}

\usepackage{etoolbox} 

\usepackage{caption}
\def\i{\mathrm{i}}
\def\e{\mathrm{e}}

\newcommand{\VarWithSub}[2]{#1_{\mathrm{#2}}}

\def\Tin{\VarWithSub{T}{in}}
\def\lengthFC{\VarWithSub{L}{fc}}
\def\loss{\Lambda}

\def\gammaIfo{\VarWithSub{\gamma}{ifo}}

\def\gammaAFC{\VarWithSub{\gamma}{afc}}
\def\K{\mathcal{K}}

\def\omegaFC{\VarWithSub{\omega}{fc}}
\def\OmegaSQL{\VarWithSub{\Omega}{SQL}}

\def\etaAFC{\VarWithSub{\eta}{afc}}
\def\rFC{\VarWithSub{r}{fc}}
\def\rAFC{\VarWithSub{r}{afc}}

\newcommand{\greenWavelength}{\SI{532}{\nano\meter}}
\newcommand{\redWavelength}{\SI{1064}{\nano\meter}}

\newcommand{\cavityLength}{\SI{16}{\meter}}

\newcommand{\armPower}{\SI{750}{\kilo\watt}}
\newcommand{\sigBW}{\SI{450}{\Hz}}

\newcommand{\ifoOmegaSQL}{\SI{63}{\hertz}}
\newcommand{\ifoSQZ}{\SI{9}{\decibel}}
\newcommand{\ifoLoss}{\SI{5}{\percent}}
\newcommand{\ROLoss}{\SI{10}{\percent}}

\newcommand{\cavityBandwidth}{\SI{90}{\hertz}}
\newcommand{\CLFfreq}{\SI{8.9}{\mega\hertz}}

\newcommand{\RLFfreq}{\SI{9.3}{\mega\hertz}}

\newcommand{\diagDetuning}{\SI{1}{\kilo\hertz}}

\newcommand{\sqzLevel}{\SI{5}{\decibel}}
\newcommand{\asqzLevel}{\SI{9}{\decibel}}
\newcommand{\gensqzLevel}{\SI{11}{\decibel}}
\newcommand{\totalLoss}{21\%}

\begin{document}

\title{Demonstration of an amplitude filter cavity at gravitational-wave frequencies}

\author{Kentaro Komori}
\email[]{kentarok@mit.edu}
\affiliation{LIGO Laboratory, Massachusetts Institute of Technology, Cambridge, Massachusetts 02139, USA}
\author{Dhruva Ganapathy}
\email[]{dhruva96@mit.edu}
\affiliation{LIGO Laboratory, Massachusetts Institute of Technology, Cambridge, Massachusetts 02139, USA}
\author{Chris Whittle}
\affiliation{LIGO Laboratory, Massachusetts Institute of Technology, Cambridge, Massachusetts 02139, USA}
\author{Lee McCuller}
\affiliation{LIGO Laboratory, Massachusetts Institute of Technology, Cambridge, Massachusetts 02139, USA}
\author{Lisa Barsotti}
\affiliation{LIGO Laboratory, Massachusetts Institute of Technology, Cambridge, Massachusetts 02139, USA}
\author{Nergis Mavalvala}
\affiliation{LIGO Laboratory, Massachusetts Institute of Technology, Cambridge, Massachusetts 02139, USA}
\author{Matthew Evans}
\affiliation{LIGO Laboratory, Massachusetts Institute of Technology, Cambridge, Massachusetts 02139, USA}

\date{\today}

\begin{abstract}
  Quantum vacuum fluctuations fundamentally limit the precision of optical measurements, such as those in gravitational-wave detectors.
  Injection of conventional squeezed vacuum can be used to reduce quantum noise in the readout quadrature, but this reduction is at the cost of increasing noise in the orthogonal quadrature.
  For detectors near the limits imposed by quantum radiation pressure noise (QRPN), both quadratures impact the measurement, and the benefits of conventional squeezing are limited.
  In this paper, we demonstrate the use of a critically-coupled \cavityLength{} optical cavity to diminish anti-squeezing at frequencies below \cavityBandwidth{} where it exacerbates QRPN, while preserving beneficial squeezing at higher frequencies. This is called an amplitude filter cavity, and it is useful for avoiding degradation of detector sensitivity at low frequencies. The attenuation from the cavity also provides technical advantages such as mitigating backscatter.  
\end{abstract}

\maketitle

\section{Introduction}

%
In recent years, precision measurements of displacements in cavity-optomechanical systems have achieved the shot-noise limits imposed by vacuum fluctuations of the optical field~\cite{Teufel2011, Chan2011, Wollman2015, Pirkkalainen2015, Peterson2016, Cripe2019}, including present-era gravitational-wave (GW) detectors~\cite{Aasi2015,Acernese2015,Aso2013}. The vacuum manifests as quantum noise in the amplitude and phase quadratures of the optical field, and may be manipulated through the technique of squeezing~\cite{Caves1981}. Phase quadrature squeezing to improve shot noise has been demonstrated in first-generation GW detectors~\cite{Abadie2011, Aasi2013, Grote2013} and in existing second-generation facilities. Injection of squeezed states enabled improvements in sensitivity corresponding to an increase in detection rate of up to 50\%~\cite{Tse2019,Acernese2019}, heralding the era of quantum-enhanced gravitational-wave astronomy.

%
The injection of phase squeezing is accompanied by a potential trade-off that is imposed by the Heisenberg uncertainty principle, whereby reducing the quantum noise in one quadrature necessitates an increase of noise in the other.
A measurement in this noisier quadrature observes a noise increase, dubbed \emph{anti-squeezing}. For GW detectors, phase quadrature squeezing is used to reduce the imprecision resulting from shot noise, so the amplitude quadrature experiences anti-squeezing. Amplitude fluctuations are responsible for variations in the radiation pressure force acting on the test masses.
Consequently, any detector observing across frequencies where the dominant quantum noise contribution transitions from quantum shot noise to quantum radiation pressure noise (QRPN) will face a trade off which limits the optimal level of squeezing.

Predominantly masked by classical noise sources, QRPN has not yet significantly degraded detector performance. Nevertheless, recent improvements in classical noises have allowed the effects of anti-squeezing to be observed~\cite{Yu2020, Acernese2020}.
In fact, the squeezing level of the vacuum state used in Advanced LIGO was optimized to minimize radiation pressure effects, with the binary neutron star detection range used as a metric of performance~\cite{Tse2019}.


The noise from anti-squeezing may be mitigated by optical loss, which attenuates optical fields---quantum or classical---while adding vacuum fluctuations. However, a broadband optical loss also degrades squeezing at high frequencies. We can preserve the benefit of squeezing at high frequencies by selectively introducing loss at low frequencies. This can be achieved by adding a resonant, critically-coupled cavity, named an amplitude filter cavity (AFC). 
This technique has been investigated theoretically~\cite{Corbitt2004}, and compared to other configurations such as the phase filtering scheme~\cite{Khalili2010}. It is experimentally demonstrated in this paper, to our knowledge, for the first time. The AFC technique is in contrast to using frequency-dependent squeezed quadrature rotation to achieve broadband squeezing. Such squeezed state rotation is implemented using a low-loss, off-resonance (detuned) filter cavity~\cite{Kimble2001, Evans2013, Miller2015}. Detuned filter cavities have been experimentally demonstrated by proof-of-principle experiments in the MHz region~\cite{Chelkowski2005}, at audio-band kHz frequencies~\cite{Oelker2016}, and now achieve sub-\SI{100}{\hertz} rotation frequencies useful for current GW detectors~\cite{McCuller2020, Zhao2020}. To perform at such low frequencies, detuned cavities need a decoherence time much longer than the storage time. This amounts to having long cavities with exceptionally low loss mirrors, and challenging length stability requirements;
 all of these requirements are relaxed for AFCs.

%

%
Here we first calculate the improvement in sensitivity of an interferometer with squeezed light injected after reflection off an amplitude filter cavity.
We then experimentally show that, using a \cavityLength{} long critically-coupled cavity, anti-squeezing below \cavityBandwidth{} is attenuated by the cavity, while squeezing above this frequency reflects unimpaired. 
The recorded data are compared with our model of the quantum noise in the system, which is additionally used to infer some difficult-to-measure system experimental parameters.
The application of the amplitude filter to GW detectors is explored in the context of the detector binary neutron star range.
Finally, we discuss the technical advantages of running a filter cavity on resonance, as required by the AFC technique.

\section{Model}
%
Here we establish a model of a GW detector employing phase quadrature
squeezing with an amplitude filter cavity. 
This model is based on the more complete description in Ref.~\cite{Whittle2020, Kwee2014},
 and is simplified to the particular application of an AFC.
We then relate it to our experimental measurement.

%
The trade off faced by GW interferometers is shown below in the noise spectrum of an ideal lossless detector, in units relative to the vacuum fluctuations:
\begin{equation}
\label{eq:totalnoisefis}
N(\Omega) = \e^{2\sigma} \K^2 + \e^{-2\sigma}.
\end{equation}
This spectrum represents purely the optical noise spectrum and not the detector sensitivity, which is a multiplicative calibration of $N(\Omega)$ into units of strain spectral density. At most frequencies, the noise scales with the squeezing level through the term $\e^{-2\sigma}$, but is limited by losses in realistic detectors to a minimum level. The $\K$ term which scales with the anti-squeezing is frequency-dependent and determined by the mirror mass, detector bandwidth $\gammaIfo$ and operating power. Those dependencies determine the scale frequency $\OmegaSQL$ at which $\K(\Omega)\simeq1$; that is, the QRPN and shot noise contribute equally to the quantum noise. In terms of this scale, the frequency dependence of $\K$, for a free mirror mass, is
\begin{equation}
\K(\Omega) = \frac{\OmegaSQL^2}{\Omega^2} \frac{\gammaIfo^2}{\gammaIfo^2 + \Omega^2}.
\end{equation}
For the current generation GW detectors, $\OmegaSQL$ spans 30Hz to 90Hz, indicating that low frequency content of astrophysical signals can be significantly degraded when squeezing is employed.

In order to allow for optical losses we extend
the noise spectrum of Eq.~\ref{eq:totalnoisefis} to account for the transmission efficiency $\eta$ between the
squeezer and interferometer;
\begin{equation}
\label{eq:lossynoise}
N(\Omega,\sigma) = \eta(\e^{2\sigma} \K^2 + \e^{-2\sigma}) + (1 - \eta)(\K^2 + 1)
\end{equation}
The first term represents squeezed vacuum that remains after the efficiency $\eta$,
 while the second term represents the un-squeezed vacuum which enters due to the optical loss.
The amplitude filter cavity serves to implement a frequency-dependent transmission efficiency
 $\eta(\Omega)$ to increase the optical loss where $\K > 1$. In the absence of squeezing, i.e $\sigma = 0$, the quantum noise is independent of optical loss and is given by $N(\Omega) = \K^2+1$.

\begin{figure*}[ht]
\centering
\includegraphics[width=\textwidth]{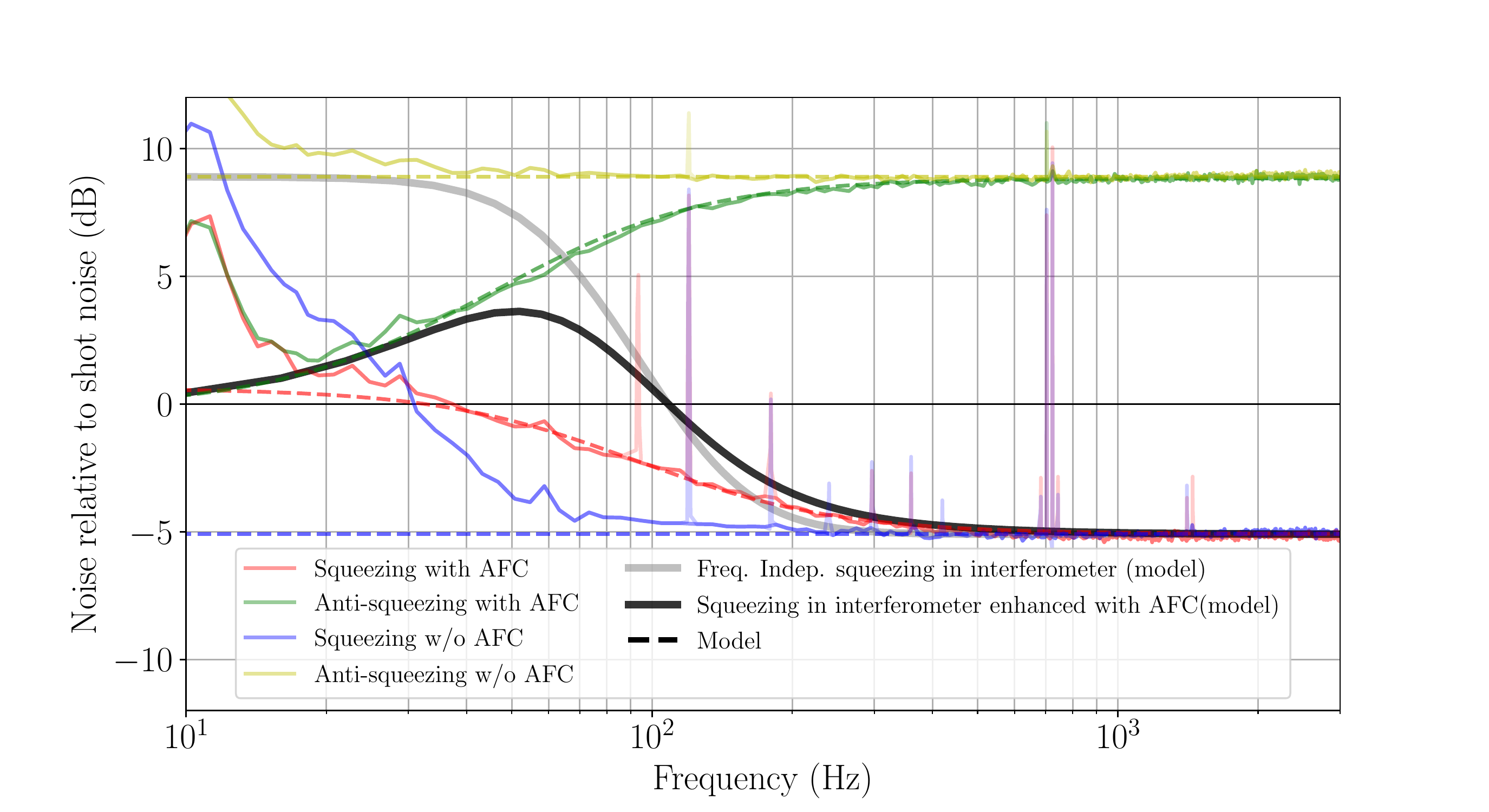}
\caption{
Demonstration of an amplitude filter cavity at gravitational-wave frequencies. The measured data (solid) are fit to the model (dashed). The noise is plotted relative to quantum shot noise (0\,dB). We demonstrate the effect of a near critically-coupled filter cavity on both squeezed (red) and anti-squeezed (green) states. We see that the curves fall toward shot noise at low frequencies as expected from a critically-coupled cavity on resonance. The slight excess with respect to shot noise at low frequencies is due to detuning noise coupled with the fact that the cavity is slightly overcoupled. For reference, we have also plotted frequency-independent squeezed (blue) and anti-squeezed (yellow) data. The noise excess over the model at lower frequencies is due to modulated back-scattered light from the homodyne detector. We observe that the amplitude filter removes some of this back-scattered light at low frequencies through loss, an effect not observed for detuned filter cavities. The acoustic peaks in the data have been excluded from frequency-bin averaging but have been included in the plot as faded traces. The black and gray solid traces are modeled noise improvements (Eqs.~\ref{eq:lossynoise} and \ref{eq:totalimpfactorhd}) when squeezing is applied to an interferometer with $\OmegaSQL$ and signal bandwidth equal to those of the test interferometer described in Table~\ref{table2}.}
\label{fig:data}
\end{figure*}

With a cavity, the complex reflectivity of the optical resonance determines the transmission efficiency.
 Using a high-finesse approximation, the reflectivity is written as
\begin{equation}
\label{eq:rfcgen}
\rFC(\Omega) = \frac{-\gamma + \lambda + \i(\Omega - \Delta \omegaFC)}{\gamma + \lambda + \i(\Omega - \Delta \omegaFC)}.
\end{equation}
Here we have defined the \emph{coupler-limited} and \emph{loss-limited bandwidths} as
\begin{equation}
\label{eq:gammalambda}
\gamma = \frac{c\Tin}{4\lengthFC},\ \lambda = \frac{c\loss}{4\lengthFC}
\end{equation}
respectively, where $\Delta\omegaFC$ is the cavity detuning with respect to the carrier frequency, $c$ is the speed of light, $\Tin$ is the transmission of the input mirror, and $\loss$ is the round-trip optical loss, including the transmission of the end mirror. $\lengthFC$ is the length of the cavity. For a critically-coupled cavity on resonance, $\gamma = \lambda$ and $\Delta \omegaFC = 0 $\,Hz, so the reflectivity is given by
\begin{equation}
\label{eq:rfcgen2}
\rAFC(\Omega) = \frac{\i\Omega}{2\gamma+ \i\Omega} = \frac{\i\Omega}{\gammaAFC + \i\Omega},
\end{equation}
where $\gammaAFC = 2\gamma$ is the bandwidth of the filter cavity. On resonance, the $+\Omega$ and $-\Omega$ sidebands encounter a symmetric response~\cite{Kwee2014} which prevents squeezed state rotation, and allows us to simplify the analysis by only considering the amplitude of the reflectivity $\etaAFC(\Omega)$. The phase of the reflectivity does not affect the quantum noise. This efficiency is given by
\begin{gather}
\label{eq:rfccomplex}
\etaAFC(\Omega) = |\rAFC|^2 =  \frac{(\Omega/\gammaAFC)^2}{1+(\Omega/\gammaAFC)^2}.
\end{gather}

To determine the merits of the amplitude filter cavity technique,
 we consider the change of the quantum noise in an interferometer with an AFC with and without squeezing. 
Assuming no losses other than those from the AFC, 
 Eq.~\ref{eq:lossynoise} gives the change of quantum noise in the detector relative to no injected squeezing
\begin{align}
\label{eq:totalimpfactor}
  I_{\text{GW}}(\Omega) &=\frac{N(\Omega,\sigma)}{N(\Omega,\sigma=0)} \\
  &=\frac{\left[ 1+ \etaAFC (\e^{2\sigma} -1) \right] \K^2 + 1 + \etaAFC ( \e^{-2\sigma}-1)}{ \K^2 +1}. \nonumber
\end{align}
$I_\text{GW}(\Omega)$ demonstrates the following two limits: at high frequencies, where $\K \ll 1$, squeezing is achieved based on the squeezing level and residual efficiencies where, ideally, $\etaAFC \simeq 1$ and $I_{\text{GW}} \simeq \e^{-2\sigma}$.
At low frequencies, where $\K\gg1$, only the radiation pressure term remains and this is where $\etaAFC \simeq 0$ and $I_{\text{GW}} \simeq 1$.

Without the interferometer, we measure the amplitude filter cavity using a balanced homodyne detector to record the noise spectrum of the squeezed field. The two limits of Eq. \ref{eq:totalimpfactor} are established by making separate measurements of the squeezing and of the anti-squeezing. The spectrum relative to the coherent-state vacuum is given by
\begin{equation}
    N_{\text{HD}}(\Omega, \pm\sigma) = \etaAFC \e^{\pm 2\sigma} + 1 - \etaAFC,
\end{equation}
where squeezing and anti-squeezing quadrature observations correspond to $-\sigma$ and $+\sigma$, respectively. Together, the two measurements allow one to construct the interferometer relative quantum noise as the weighted average of the homodyne measurements:
\begin{equation}
\label{eq:totalimpfactorhd}
    I_{\text{GW}}(\Omega, \sigma) = \frac{N_{\text{HD}}(\Omega, +\sigma)\K^2 + N_{\text{HD}}(\Omega, -\sigma)}{\K^2 + 1}.
\end{equation}
Fig.~\ref{fig:data} shows the modeled interferometer quantum noise and its subsequent improvement upon the introduction of an amplitude filter cavity. We also show the results of these two quadrature measurements, the experimental setup for which is described in the next section.

\begin{figure*}[]
\centering
\includegraphics[width=0.85\textwidth]{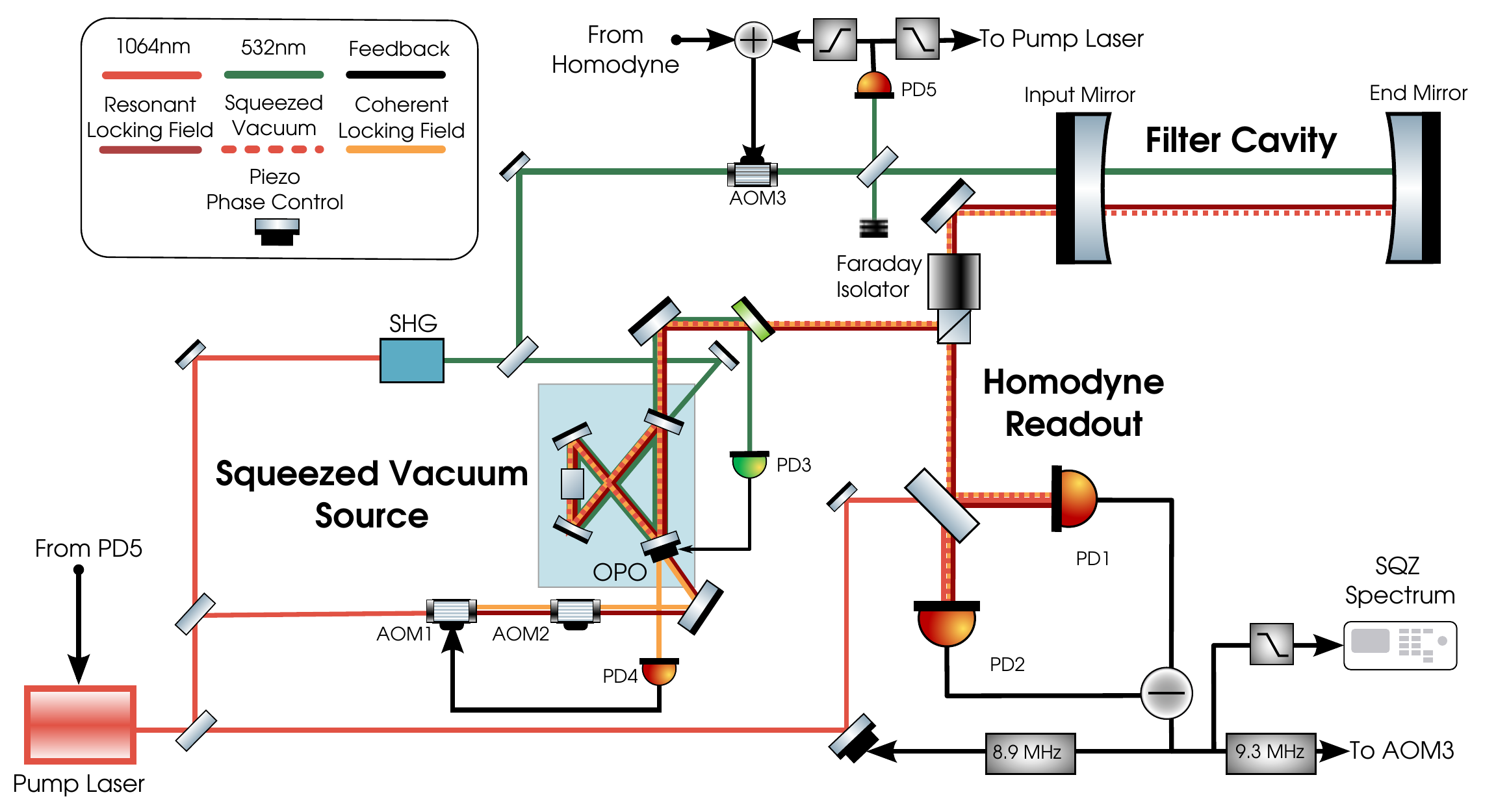}
\caption{
Experimental layout of an amplitude filter cavity for gravitational-wave detectors. The output of a \redWavelength{} laser is split into three parts which are:  i) up-converted to \greenWavelength{} by second harmonic generation (SHG) and used as the pump of the Optical Parametric Oscillator (OPO), ii) used for the local oscillator for homodyne detection, and iii) passed through AOM1 and AOM2 to generate the coherent and resonant locking fields for controlling the squeezing quadrature and filter cavity detuning respectively. Squeezed vacuum is generated by an in-vacuum doubly resonant OPO and undergoes frequency dependent attenuation on reflection from the filter cavity, locked to resonance for the carrier field. A Faraday isolator then re-directs the squeezed light to the homodyne readout for detection. 
}   
\label{fig:setup}
\end{figure*}

\section{Experimental setup}

%
We demonstrate an amplitude filter using a \cavityLength{} critically-coupled cavity. A schematic of our experiment is shown in Fig.~\ref{fig:setup} and the key parameters of our setup are listed in Table~\ref{table1}. This setup is nearly identical to our previous demonstration of frequency-dependent squeezing \cite{McCuller2020} except for the cavity loss. We increased this loss to achieve a critically-coupled cavity by changing the beam spot position off the optimal point for the cavity optics. A portion of the \redWavelength{} pump laser is picked off and up-converted to \greenWavelength{} via second harmonic generation (SHG). A portion of the \greenWavelength{} light is double-passed through an acousto-optic modulator (AOM) and is used to coarsely lock the \redWavelength{} carrier to the filter cavity resonance with Pound-Drever-Hall sensing~\cite{Black2001}. The remaining \greenWavelength{} light pumps an optical parametric oscillator (OPO) to generate frequency-independent squeezed vacuum. The generated squeezed vacuum is reflected off of the filter cavity and is subsequently measured with a balanced homodyne readout (PD1 and PD2 in Fig.~\ref{fig:setup})
 employing a $\sim$\SI{0.24}{\milli\watt} local oscillator (LO).

\begin{table}[]
\centering
\caption{Parameters for our amplitude filter cavity experiment. Entries marked by an asterisk were determined by fitting to recorded data. In all cases fitting produced values consistent with independent measurements and their uncertainties.}
\begin{tabular}{cc}
  \hline
  Parameter                          & Value \\
  \hline                                                   
  \hline
  Filter cavity length               & \SI{16.0611 \pm 0.0002}{m}\\
  Filter cavity storage time      & \SI{2.1 \pm 0.1}{ms}\\
  OPO nonlinear gain$^*$             & \SI{4.6 \pm 0.1}{}      \\
  OPO escape efficiency              & \SI{98 \pm 1}{\percent}   \\
  Propagation loss$^*$               & \SI{15 \pm 1}{\percent}    \\
  Homodyne visibility                & \SI{96.8 \pm 0.7}{\percent}                   \\
  Photodiode quantum efficiency      & \SI{99 \pm 1}{\percent}         \\
  Input mirror transmission       & \SI{51.5 }{ppm}                   \\
  Filter cavity round-trip loss       & \SI{46 }{ppm}                   \\
 Freq. indep. phase noise (RMS)$^*$ & \SI{10 \pm 5}{mrad}   \\
 Detuning fluctuation (RMS)$^*$ & \SI{12.5}{Hz}     \\
Filter cavity mode matching        & \SI{94 \pm 1}{\percent}      \\
\hline
\end{tabular}
\label{table1}
\end{table}

Two acousto-optic modulators (AOM1 and AOM2) shift a pickoff of \redWavelength{} light by \CLFfreq{} and \RLFfreq{}. The squeezing angle is controlled using the \CLFfreq{}-shifted light, known as the coherent locking field (CLF). The beatnote of the two CLF sidebands on reflection from the OPO and the beatnote between the CLF and LO at the homodyne detector provide the error signals for the squeezing angle. The \RLFfreq{}-shifted light is called the resonant locking field (RLF) and provides the second layer of control over the cavity detuning. The detuning error signal is provided by the beatnote between the RLF and LO, also measured at the homodyne
(see \cite{McCuller2020} for more information).


\section{Squeezing spectra}
Fig.~\ref{fig:data} shows the measured squeezing spectra normalized to quantum shot noise. First, the noise spectrum is measured in the absence of squeezed vacuum with the local oscillator to determine the shot noise (\SI{0}{\decibel}) reference. Next, frequency-independent squeezing and anti-squeezing data are taken to estimate the squeezing level generated ($\e^{\pm 2\sigma}$) and injection/readout optical loss. We measure \sqzLevel{} squeezing and \asqzLevel{} anti-squeezing respectively, which implies \gensqzLevel{} generated squeezing and \totalLoss{} total optical loss. For the above measurements, we hold the filter cavity far from resonance with respect to the carrier using the \greenWavelength{} light to avoid any resonant effects. Lastly, we repeat the squeezing and anti-squeezing measurements with the filter cavity resonant for the carrier. Our measured data (solid traces) fit our model (dotted traces) well. Except for the detuning and squeezer angle which are inferred from the fit, all other parameters of the setup are measured independently (see Table~\ref{table1}). Fig.~\ref{fig:data} also shows the modeled interferometer quantum noise and its subsequent improvement upon the introduction of an AFC. The red trace, corresponding to the squeezing with the AFC, slightly exceeds shot noise at low frequencies due to a combination of detuning fluctuation and the fact that the cavity is slightly overcoupled. 
\begin{figure}[]
\centering
\includegraphics[width=\hsize]{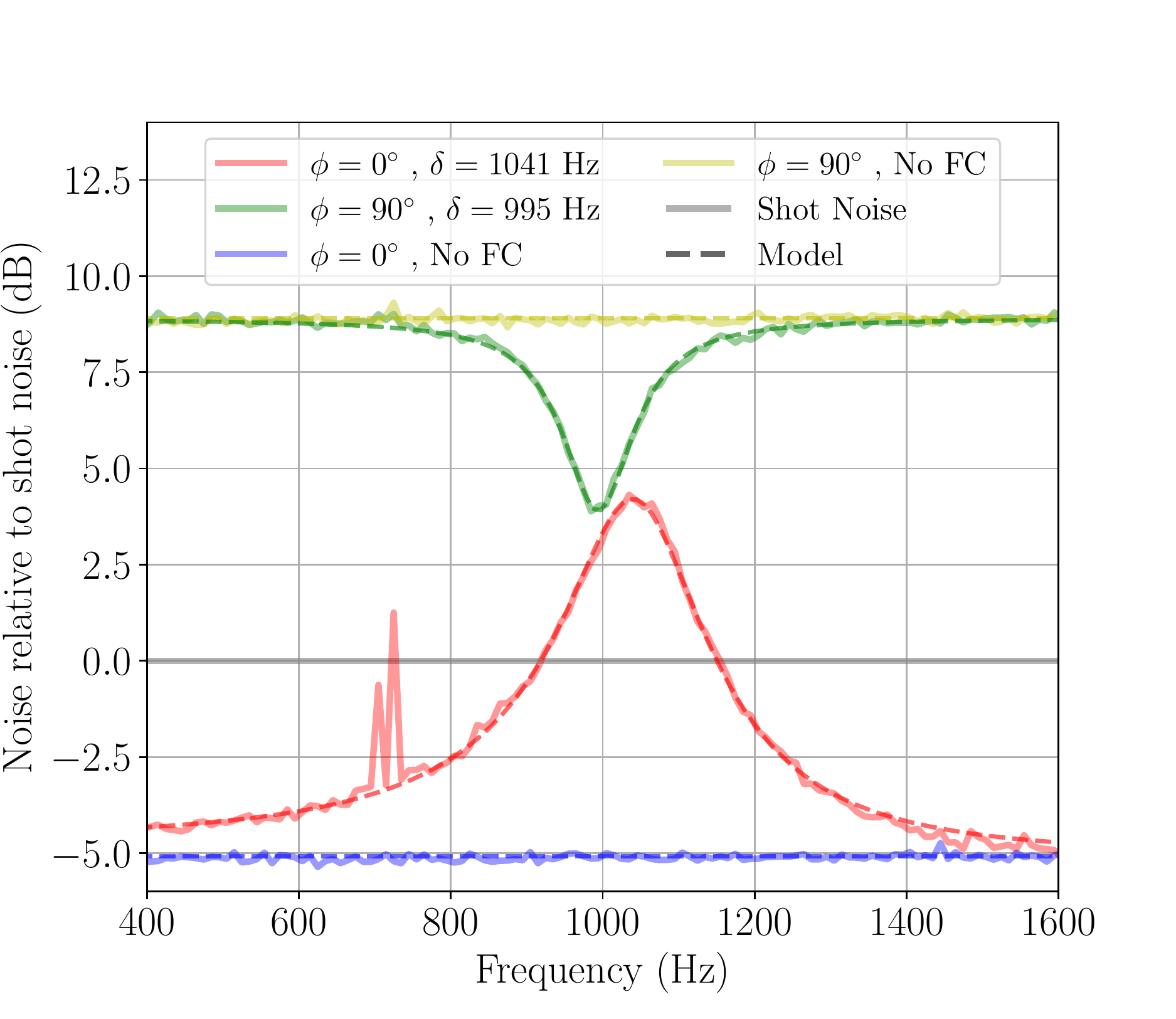}
\caption{
Squeezing spectra with the filter cavity detuned \diagDetuning{}. The detuning fluctuation of the filter cavity and the propagation loss of the squeezed vacuum are inferred by fitting our model to the measured data. The other parameters used in the model were measured directly. Acoustic peaks have been omitted for fitting to the quantum noise.
}
\label{fig:detuned}
\end{figure}

Some technical noise artifacts are visible in our measured spectra.
The peak around \SI{10}{\hertz} is due to a mechanical resonance in the optics table. Harmonics of the power supply and acoustic peaks appear above \SI{100}{\hertz}. There is a broad excess below several tens of Hz that is attributed to back-scattered local oscillator light reflecting off the homodyne optics. This light propagates through the squeezed vacuum path, leaking through the Faraday isolator before being reflected from the filter cavity to return to the homodyne detector along with the squeezed vacuum field. We note that the back-scatter noise apparent in the spectra taken without the filter cavity (e.g., the blue curve Fig.~\ref{fig:data}) is diminished in the spectra measured with the AFC due to the attenuation of the cavity. 
%

We additionally measured spectra with the filter cavity detuned at around \diagDetuning{} (See Fig. \ref{fig:detuned}) to characterise the detuning fluctuation, arising from residual cavity length noise and frequency noise on the laser,  which is difficult to measure independently. The cavity pole and mode-matching into the cavity were determined from independent measurements. We infer the detuning fluctuation in our cavity to be $12 \pm 4$\,Hz, implying a residual length noise of $0.7\pm 0.2$\,pm. This measurement is performed at high frequency to avoid bias from the back-scatter noise present at low frequencies.

\section{Application to interferometer}

\subsection{Improvement in detector range}
\begin{figure*}
\centering
\includegraphics[width=0.45\linewidth]{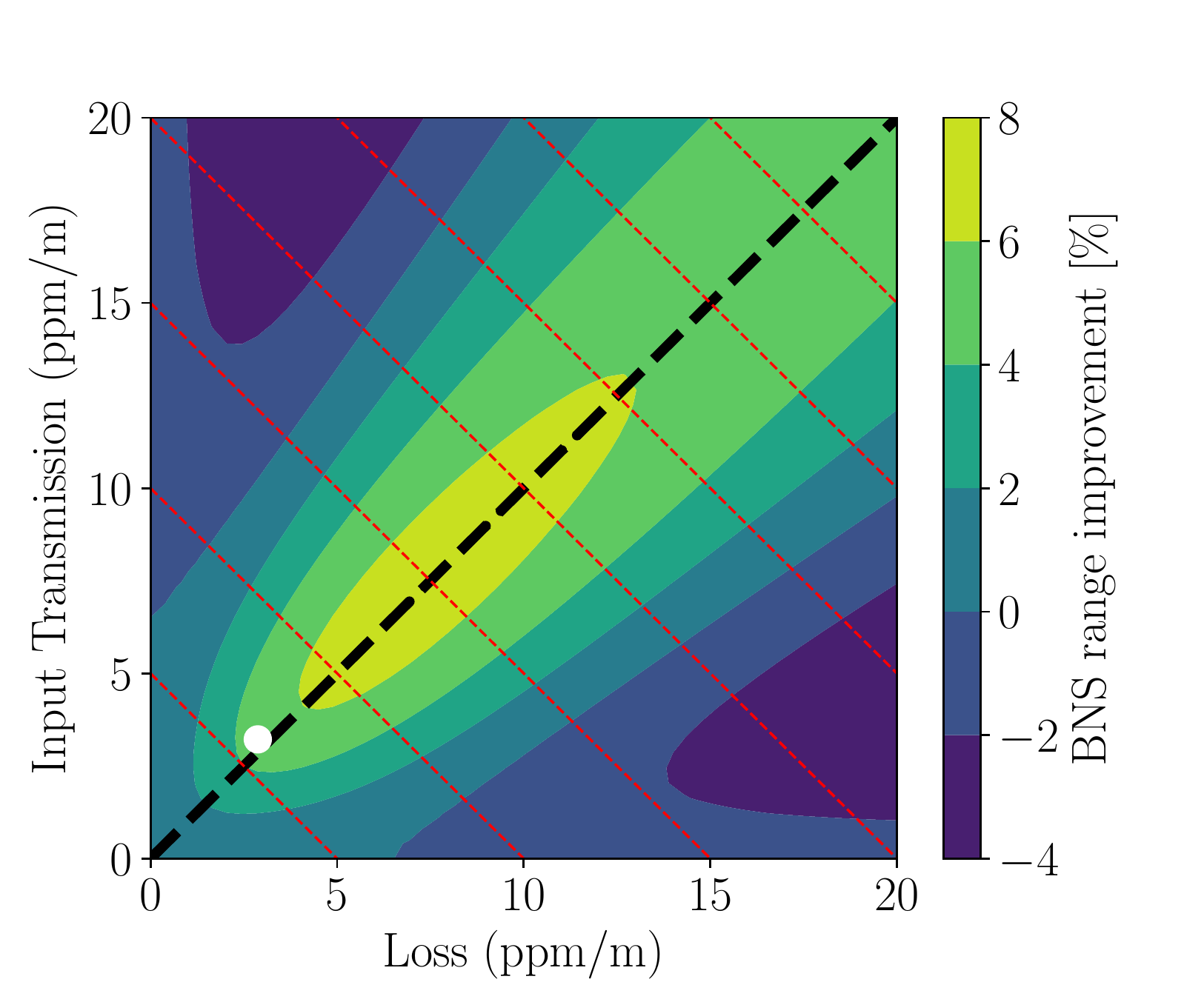}
\includegraphics[width=0.45\linewidth]{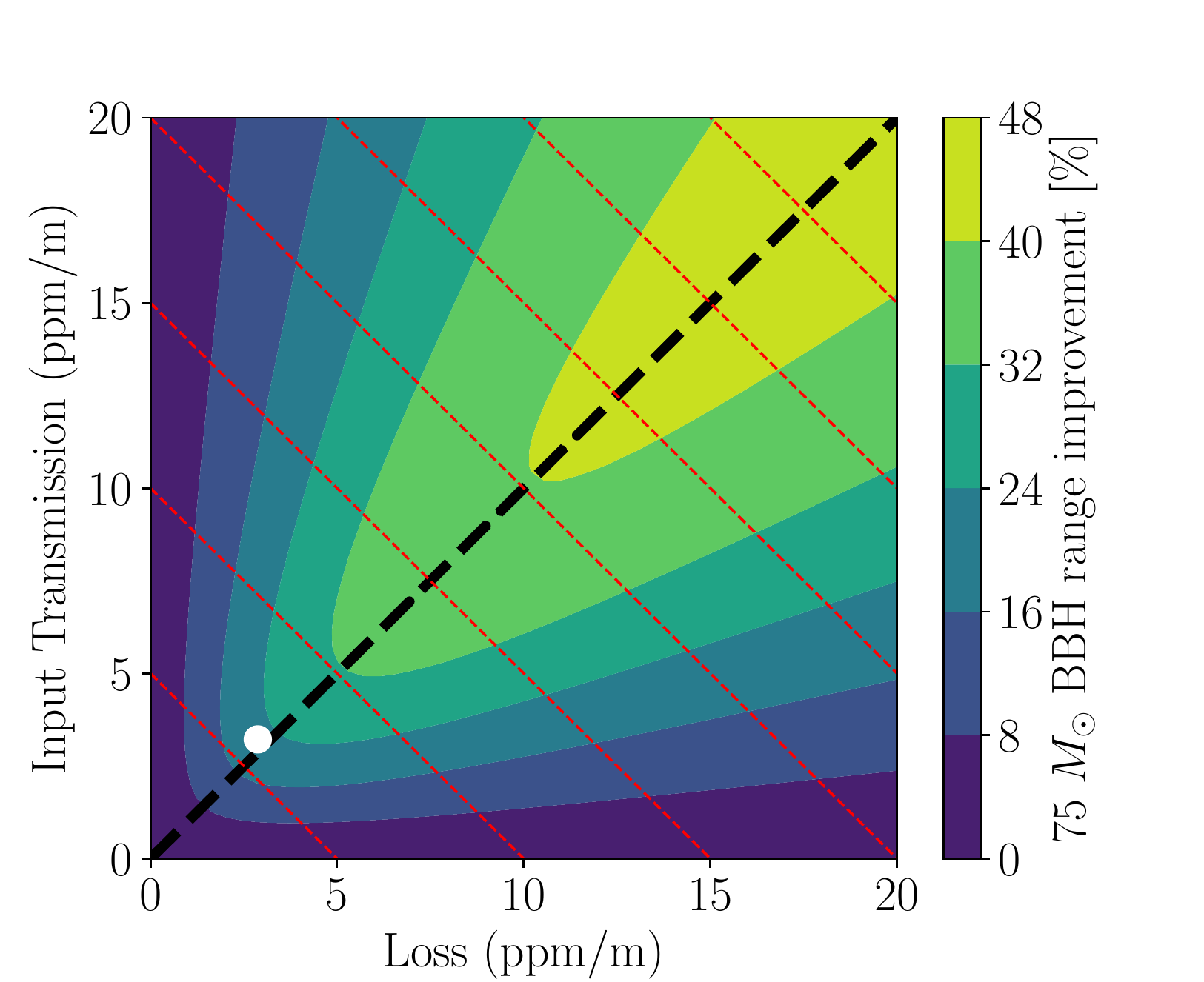}
\caption{
Enhancement in 1.4-1.4$M_{\odot}$ binary neutron star (left) and 75-75$M_{\odot}$ binary black hole (right) inspiral detection range for filter cavities with detuning $\Delta\omega_{\text{fc}}$  set to 0\,Hz. The colorbar represents the relative increase/decrease in binary inspiral range when the AFC is applied versus a squeezing-enhanced interferometer (parameters listed in Table~\ref{table2}) without an AFC. The black dashed line corresponds to critically-coupled AFCs and the white marker corresponds to the setup that we have demonstrated in this paper, the parameters of which are listed in Table~\ref{table1}. Filter cavity bandwidths are constant along the red dashed lines. It is evident that, for a given bandwidth, the range enhancement is maximum when the detector is a critically coupled AFC.
}
\label{fig:range}
\end{figure*}

In this section, we discuss advantages of the amplitude filter cavity when integrated into a gravitational-wave detector. Fig.~\ref{fig:range} shows the increase/decrease in the binary inspiral detection ranges when a filter cavity with $\Delta\omega_{\text{fc}} = 0$\,Hz is applied to an interferometer with parameters listed in Table~\ref{table2}. We see that, for a resonant cavity with a given bandwidth, the improvement in range is maximum when it is critically coupled. Additionally, the improvement in range also increases with bandwidth to a point. This is expected; anti-squeezing, which enhances QRPN caused by interferometer back-action, is destroyed by greater cavity losses. For binary neutron stars however, increasing the AFC bandwidth can become detrimental to the range by degrading squeezing at frequencies which are not affected by QRPN. Heavier binary inspirals, such as those of two $75M_{\odot}$ black holes, merge at lower frequencies and are completely dominated by QRPN, resulting in the detection range continuing to improve with increasing filter cavity bandwidth. 
\begin{table}
\centering
\caption{Assumed parameters of an interferometeric GW detector.}
\begin{tabular}{lc}
\hline
Parameter & Value \\
\hline \hline
Arm power & \armPower{} \\
Signal bandwidth & \sigBW{} \\
Scale $\OmegaSQL$ & \ifoOmegaSQL{} \\
Classical noises & Thermal noise \cite{Evans2013} \\
\hline
Injected squeezing & \ifoSQZ{} \\
Injection loss & \ifoLoss{}\\
Readout loss & \ROLoss{}\\
\hline
\end{tabular}
\label{table2}
\end{table}

\subsection{Technical advantages of an amplitude filter cavity}

The AFC configuration has a number of technical advantages. Firstly, it relies on larger optical losses in the cavity for critical coupling and consequently, considerably shorter cavities are sufficient to achieve the desired bandwidth. For optomechanical detectors with limitations on the length or losses of a filter cavity, an amplitude filter can therefore offer low-frequency improvements with more forgiving requirements with respect to a detuned filter cavity. 

Secondly, AFCs attenuate classical as well as quantum noise, relaxing scatter noise requirements on the optics relaying beams between the cavity, squeezed-state source, and interferometer. It is possible to mitigate scattered light noise by using optical isolation elements, but this comes at the cost of increasing broadband propagation loss on the squeezed vacuum path. As we have observed in our data, an amplitude filter cavity partially solves the back-scatter problem by destroying back-scattered light within its bandwidth through loss. 

Finally, AFCs have relaxed detuning noise requirements relative to detuned filter cavities as they do not need to be precisely held on resonance to achieve attenuation.
Detuning fluctuation causes some of the squeezed vacuum to rotate into the orthogonal anti-squeezed quadrature, thereby increasing noise. In order to compare the impact of the detuning fluctuation to the filter cavities, we calculate $\partial \alpha_\mathrm{p} / \partial \Delta \omega_\mathrm{fc}$ (using Eq. 4 from \cite{Whittle2020}) at $\Omega = \gamma$ for a loss-less detuned filter cavity where $(\lambda, \Delta \omega_\mathrm{fc})=(0, \gamma)$, and that for an AFC where $(\lambda, \Delta \omega_\mathrm{fc})=(\gamma, 0)$. The phase shift upon reflection from the filter cavity is 3-fold less sensitive to detuning fluctuations in the AFC than in the detuned filter cavity. Moreover, for an AFC, the rotation of the vacuum into another quadrature is partly negated by the fact that some of this excess noise is degraded to shot noise levels by cavity losses.
Numerical simulations of the quantum noise at \SI{100}{\hertz} show that detuning fluctuations degrade noise by \SI{2}{\decibel} with the detuned filter cavity, but only \SI{0.06}{\decibel} with the AFC, for the parameters from Table~\ref{table2} and an RMS detuning of \SI{10}{\hertz}.  


\section{Conclusion}

Quantum noise can be reduced through the injection of squeezed vacuum into the anti-symmetric port of a gravitational-wave detector. Presently, frequency-independent squeezing provides an improvement in detector sensitivity with little penalty. However, as low-frequency classical noise sources are reduced, degradation by anti-squeezing will increasingly limit detector performance. In future designs, critically-coupled, resonant filter cavities are one possible candidate to alleviate this degradation, mitigating the deleterious effects of squeezing at low frequencies while retaining high-frequency improvements. In this paper, we experimentally demonstrate a \cavityLength{} amplitude filter cavity. Further, we computationally find improvements in detection ranges of binary neutron star and binary black hole events through use of an amplitude filter. We demonstrate and discuss the backscatter suppression offered by this scheme. We conclude that using an amplitude filter cavity is a reasonable alternative to detuned filter cavities to improve performance of gravitational-wave detectors.

\section*{Acknowledgements}
LIGO was constructed by the California Institute
 of Technology and Massachusetts Institute of Technology with funding from the
 National Science Foundation, and operates under Cooperative Agreement No. PHY-1764464. Advanced LIGO was built under Grant No. PHY-0823459. KK is supported by JSPS Overseas Research Fellowship.
 
Over the years, a number of people have contributed to the experimental set-up used  for this measurement, in particular: Maggie Tse, Alvaro Fernandez-Galiana, Peter Fritschel, Myron MacInnis, Fabrice Matichard, Ken Mason, Rich Mittleman, Haocun Yu, Mike Zucker, John Miller, Georgia Mansell, Evan Hall and Vivishek Sudhir.

We would additionally like to thank Thomas Corbitt, Aaron Buikema and Eugene Knyazev for useful comments.

This paper has LIGO Document Number LIGO-P2000241.

\bibliography{paper}
\bibliographystyle{apsrev.bst}
\end{document}